\UseRawInputEncoding
\documentclass[journal=jpclcd,manuscript=letter]{achemso}

\usepackage[version=3]{mhchem} 
\usepackage{graphicx} 
\usepackage{subcaption} 
\usepackage{hyperref}
\usepackage{color}
\usepackage{tabularx}
\usepackage{physics}
\usepackage[utf8]{inputenc}
\usepackage[T1]{fontenc}
\setkeys{acs}{maxauthors=999}

\mathchardef\mhyphen="2D

\author{Rugwed Lokhande}
\affiliation{Department of Chemistry, University of Florida, Gainesville, FL 32603, USA}
\alsoaffiliation{Quantum Theory Project, University of Florida, Gainesville, FL 32603, USA}

\author{Krisztina Zsigmond}
\affiliation{Department of Chemistry, University of Florida, Gainesville, FL 32603, USA}
\alsoaffiliation{Quantum Theory Project, University of Florida, Gainesville, FL 32603, USA}
\author{Paul W. Ayers}
\affiliation{Department of Chemistry and Chemical Biology, McMaster University, Hamilton Ontario L8S 4M1, Canada}
\email{ayers@mcmaster.ca}
\author{Ramón Alain Miranda-Quintana}
\affiliation{Department of Chemistry, University of Florida, Gainesville, FL 32603, USA}
\alsoaffiliation{Quantum Theory Project, University of Florida, Gainesville, FL 32603, USA}
\email{quintana@chem.ufl.edu}

\title[Second-Derivative-Corrected FANPT for Robust Continuation of Nonlinear Wavefunction Equations]
  {Second-Derivative-Corrected FANPT for Robust Continuation of Nonlinear Wavefunction Equations}

\abbreviations{}

\begin{document}







\begin{abstract}
We present a second-derivative-corrected extension of the Flexible Ansatz for N-body Perturbation Theory (FANPT) for solving nonlinear Flexible Ansatz for N-body Configuration Interaction (FANCI) wavefunction equations. The original quasilinear FANPT approximation neglects second- and higher-order derivatives of the determinant overlap with respect to wavefunction parameters. In this work, we retain the overlap Hessian and neglect only third- and higher-order parameter derivatives, thereby including the leading nonlinear response of the wavefunction ansatz while preserving the same response-matrix structure used in the original FANPT formulation. The resulting additional terms enter only through the constant vector of the response equations and are implemented for coupled-cluster wavefunctions in the FanPy/FANCI framework.

The new approximation is tested on the Lithium Hydride molecule and the insertion of Beryllium into \ce{H2} using seniority-restricted coupled-cluster wavefunctions in the STO-6G basis. The main advantage of the second-derivative correction is that it provides a better initial guess for solving the projected FANCI equations at the next point along the adiabatic connection. This improvement is diagnosed by comparing the FANPT-propagated parameters with the independently optimized FANCI parameters at the same value of $\lambda$. For nonlinear coupled-cluster ans\"{a}tze, especially when larger $\lambda$-steps are used, the corrected approximation reduces the same-$\lambda$ parameter deviations and suppresses large parameter-space excursions. These results show that the leading nonlinear overlap correction improves the reliability of FANPT as a continuation strategy for nonlinear wavefunction equations.
\end{abstract}

\section{Introduction}

The accurate solution of the electronic Schr\"odinger equation remains one of the central challenges of quantum chemistry. 
Although full configuration interaction (FCI) provides the exact solution within a finite one-particle basis, its exponential scaling restricts its use to small molecular systems. \cite{sa1996, ht2013}
Practical electronic-structure methods therefore rely on approximate wavefunction parametrizations that retain the most important correlation effects while remaining computationally tractable. 
Configuration interaction (CI)\cite{sa1996, ht2013, spe1980, rb1981, rb1980, rb1987, lpf2022, ucj2016, efa2016, lra_2026,knowlesDeterminantBasedFull1989,
knowlesNewDeterminantbasedFull1984,
knowlesVeryLargeFull1989}, coupled cluster (CC) \cite{br2007, br2009, brj2010, lra_2026_1}, geminal-based wave functions\cite{hj1971,bk2022,xf2007,gs2004,rv2002,md2001,md2000,ri1999,vs1996,hs1974,cj1972,pj1972,ss1971,sd1971,hs1970,sd1970,sd1969,pf1968,ca1965,fs1965,vd2017, gpb2924,surjanIntroductionTheoryGeminals1999,
surjanStronglyOrthogonalGeminals2012,
tecmerAssessingAccuracyNew2014a,
johnsonRichardsonGaudinMeanfield2020,richerGraphicalApproachInterpreting2025}, tensor-network states \cite{vf2009,vg2008,vg2007,wr1992,ds2020,
martiDensityMatrixRenormalization2010,changlaniApproximatingStronglyCorrelated2009a,martiCompletegraphTensorNetwork2010a}, and related ansatzes \cite{ws2016,su2011,rs1995,or2014,cj2010,vf2008,vf2021,mi2007} all represent different compromises between accuracy, compactness, and computational cost. 
A common difficulty shared by many of these approaches is that the resulting amplitude or parameter equations are nonlinear and may become difficult to solve, particularly in regimes dominated by strong or static correlation.

The Flexible Ansatz for N-body Configuration Interaction (FANCI) provides a general framework for expressing many different wavefunction forms through their overlaps with Slater determinants. \cite{ap2021} 
In this formulation, the wavefunction is written as
\begin{equation}
    |\Psi(\mathbf{P})\rangle = \sum_{\mathbf{m}\in S} f_{\mathbf{m}}(\mathbf{P}) |\mathbf{m}\rangle ,
\end{equation}
where $|\mathbf{m}\rangle$ is a Slater determinant in a chosen determinant space $S$, $\mathbf{P}$ is the set of wavefunction parameters, and 
$f_{\mathbf{m}}(\mathbf{P}) = \langle \mathbf{m} | \Psi(\mathbf{P}) \rangle$ is the overlap between the ansatz and the determinant. 
The wavefunction parameters and energy are then obtained by solving the projected Schr\"odinger equation,
\begin{equation}
    \sum_{\mathbf{m}\in S} 
    f_{\mathbf{m}}(\mathbf{P})
    \langle \mathbf{n} | \hat{H} | \mathbf{m} \rangle
    =
    E f_{\mathbf{n}}(\mathbf{P}),
    \qquad \mathbf{n}\in S_{\mathrm{proj}} .
\end{equation}
Because this formulation depends only on the overlap function, it can accommodate conventional CI, coupled cluster, geminal wave functions, and other nonlinear ans\"{a}tze within a unified projected-equation formalism.

The Flexible Ansatz for N-body Perturbation Theory (FANPT) was introduced as a perturbative strategy for guiding the solution of these nonlinear FANCI equations. \cite{ap2024}
In FANPT, the physical Hamiltonian is connected to a simpler zeroth-order Hamiltonian (usually the Fock operator) through an adiabatic connection,
\begin{equation}
    \hat{H}(\lambda) = \hat{F} + \lambda \hat{V},
\end{equation}
where $\lambda$ evolves from $0$ to $1$. 
At each value of $\lambda$, FANPT estimates the changes in the wavefunction parameters and energy through a Taylor expansion in $\lambda$. 
These estimates can then be used either as perturbative approximations \cite{lj1955,ph1941,co2000,sf2000,sd1998,mm1996,hm1989,br1981} or, more importantly, as improved initial guesses for solving the full projected FANCI equations at the next point along the adiabatic connection. This is analogous in spirit to using Møller--Plesset perturbation theory to generate initial coupled-cluster amplitudes.\cite{gd2000,hm1997,jp1996,aj1996,rm1988,ni1998,mr1991,kg2010,sc2010}
This makes FANPT particularly useful as a robust continuation method for difficult nonlinear wavefunction equations.

In the original FANPT formulation, the response equations were developed using what we refer to here as the quasilinear approximation order (qao). In this approximation, second and higher derivatives of the overlap function with respect to the wavefunction parameters are neglected (qao=2):
\begin{equation}
\frac{\partial^r f_{\mathbf{m}}(\mathbf{P})}
{\partial p_{k_1}\partial p_{k_2}\cdots \partial p_{k_r}}
\approx 0,
\qquad r \geq 2 .
\end{equation}
This approximation greatly simplifies the FANPT hierarchy in terms of $n^{th}$ order corrections. Because the overlap is treated as locally linear in the wavefunction parameters, the same FANPT response matrix can be used at every perturbation order, and higher-order corrections can be generated without constructing a new linear system at each order. For linear wavefunction parametrizations, such as CI expansions, this approximation is exact because the determinant overlaps are linear functions of the CI coefficients. For nonlinear ans\"{a}tze, such as coupled cluster and geminal-based wave functions, however, the neglected overlap derivatives are generally nonzero. The success of the original FANPT implementation therefore shows that the quasilinear approximation can be effective in practice, but it also raises a natural question: can the FANPT continuation path be improved by retaining the leading nonlinear overlap response?

In this work, we develop and implement a second derivative corrected FANPT approximation. Instead of neglecting all second and higher wavefunction-parameter derivatives, we retain the overlap Hessian and neglect only third and higher wavefunction-parameter derivatives (qao=3):
\begin{equation}
\frac{\partial^r f_{\mathbf{m}}(\mathbf{P})}
{\partial p_{k_1}\partial p_{k_2}\cdots \partial p_{k_r}}
\approx 0,
\qquad r \geq 3 .
\end{equation}
Thus, the present approximation includes terms involving
\begin{equation}
\frac{\partial^2 f_{\mathbf{m}}(\mathbf{P})}
{\partial p_k \partial p_l},
\end{equation}
while preserving much of the computational structure of the original quasilinear scheme.

The practical implementation of this second derivative corrected approximation requires one central new ingredient: the second derivative of the overlap vector with respect to the active wavefunction parameters. For coupled cluster wave functions, this corresponds to differentiating the product structure of excitation amplitudes that contributes to each determinant overlap. Once the overlap Hessian is available, the additional residual derivative tensors can be constructed within the existing FANCI/FANPT derivative infrastructure. In this way, the new approximation extends the original quasilinear FANPT method by including the leading nonlinear response of the wavefunction ansatz while retaining the same general continuation framework.

The goal of this manuscript is to present the mathematical formulation, implementation, and numerical behavior of this second derivative corrected FANPT approximation. We first derive the modified second- and third-order corrections FANPT equations under the new approximation. We then describe the implementation of the overlap Hessian and the associated residual derivative tensors in the FanPy/FANCI framework. Finally, we assess the effect of the new terms for representative nonlinear wavefunction ansatzes. By comparing the original quasilinear FANPT approximation with the present second-derivative-corrected approximation, we clarify when the neglected nonlinear overlap response is important and when the simpler quasilinear treatment is already sufficient.

\section{Theory}

\subsection{FANCI residual equations and the FANPT adiabatic connection}

In the FANCI framework, the wavefunction is represented through its overlaps with
Slater determinants,
\begin{equation}
    |\Psi(\mathbf{P})\rangle
    =
    \sum_{m \in S} f_m(\mathbf{P}) |m\rangle ,
\end{equation}
where \(S\) is the determinant space used to define the wavefunction, \(\mathbf{P}\)
is the vector of wavefunction parameters, and
\begin{equation}
    f_m(\mathbf{P}) = \langle m | \Psi(\mathbf{P}) \rangle
\end{equation}
is the overlap between the ansatz and determinant \(|m\rangle\). The wavefunction
parameters and the energy are determined by solving the projected Schr\"odinger
equation,
\begin{equation}
    \sum_{m \in S}
    f_m(\mathbf{P})
    \langle n | \hat{H} | m \rangle
    -
    E f_n(\mathbf{P})
    =
    0,
    \qquad n \in S_{\mathrm{proj}},
\end{equation}
where \(S_{\mathrm{proj}}\) is the projection space.

In FANPT, the physical Hamiltonian is reached through the adiabatic connection
\begin{equation}
    \hat{H}(\lambda)
    =
    \hat{F}
    +
    \lambda \hat{V},
\end{equation}
where \(\lambda=0\) corresponds to the zeroth-order Hamiltonian and \(\lambda=1\)
corresponds to the physical Hamiltonian. The corresponding projected residual is
defined as
\begin{equation}
    G^{(0)}_n(\lambda,E_\lambda,\mathbf{P}_\lambda)
    =
    \sum_{m \in S}
    f_m(\mathbf{P}_\lambda)
    \langle n | \hat{F}+\lambda \hat{V} | m \rangle
    -
    E_\lambda f_n(\mathbf{P}_\lambda).
    \label{eq:fanpt_residual}
\end{equation}
The FANPT equations for $r^{th}$ order are obtained by enforcing
\begin{equation}
    G^{(r)}_n(\lambda,E_\lambda,\mathbf{P}_\lambda)=0,
    \qquad n \in S_{\mathrm{proj}},
\end{equation}
and taking the linear (i.e., total) differential of this equation with respect to $\lambda, E_\lambda \hspace{0.5em} \mathrm{and} \hspace{0.5em} \mathbf{P}_\lambda$.

The energy and wavefunction parameters at a nearby value of the perturbation
parameter are expanded as
\begin{equation}
    E_{\lambda+\Delta\lambda}
    =
    E_\lambda
    +
    \sum_{r=1}^{N}
    \frac{1}{r!}
    \frac{d^r E_\lambda}{d\lambda^r}
    (\Delta\lambda)^r,
    \label{eq:energy_taylor}
\end{equation}
and
\begin{equation}
    p_{k,\lambda+\Delta\lambda}
    =
    p_{k,\lambda}
    +
    \sum_{r=1}^{N}
    \frac{1}{r!}
    \frac{d^r p_{k,\lambda}}{d\lambda^r}
    (\Delta\lambda)^r.
    \label{eq:param_taylor}
\end{equation}
For compactness, we define
\begin{equation}
    E^{(r)} \equiv \frac{d^r E_\lambda}{d\lambda^r},
    \qquad
    p_k^{(r)} \equiv \frac{d^r p_{k,\lambda}}{d\lambda^r}.
\end{equation}

We also introduce the following shorthand notation for derivatives of the residual:
\begin{equation}
    G_{n,k}
    \equiv
    \frac{\partial G_n}{\partial p_k},
    \qquad
    G_{n,E}
    \equiv
    \frac{\partial G_n}{\partial E},
    \qquad
    G_{n,\lambda}
    \equiv
    \frac{\partial G_n}{\partial \lambda},
\end{equation}
\begin{equation}
    G_{n,kl}
    \equiv
    \frac{\partial^2 G_n}{\partial p_k \partial p_l},
    \qquad
    G_{n,kE}
    \equiv
    \frac{\partial^2 G_n}{\partial p_k \partial E},
    \qquad
    G_{n,k\lambda}
    \equiv
    \frac{\partial^2 G_n}{\partial p_k \partial \lambda},
\end{equation}
and
\begin{equation}
    G_{n,klE}
    \equiv
    \frac{\partial^3 G_n}{\partial p_k \partial p_l \partial E},
    \qquad
    G_{n,kl\lambda}
    \equiv
    \frac{\partial^3 G_n}{\partial p_k \partial p_l \partial \lambda}.
\end{equation}
All derivatives are evaluated at the current value
\((\lambda,E_\lambda,\mathbf{P}_\lambda)\) unless otherwise stated.

\subsection{Residual derivative tensors generated by the overlap Hessian}

Using the residual definition in Eq.~\eqref{eq:fanpt_residual}, the derivative of
\(G_n\) with respect to a wavefunction parameter is
\begin{equation}
    G_{n,k}
    =
    \sum_{m \in S}
    \frac{\partial f_m}{\partial p_k}
    \langle n | \hat{F}+\lambda \hat{V} | m \rangle
    -
    E_\lambda
    \frac{\partial f_n}{\partial p_k}.
    \label{eq:first_param_deriv}
\end{equation}

The original quasilinear approximation keeps only derivatives involving at most
one wavefunction parameter derivative. In contrast, the present approximation
requires the second derivative of the residual with respect to two wavefunction
parameters,
\begin{equation}
    G_{n,kl}
    =
    \sum_{m \in S}
    \frac{\partial^2 f_m}{\partial p_k \partial p_l}
    \langle n | \hat{F}+\lambda \hat{V} | m \rangle
    -
    E_\lambda
    \frac{\partial^2 f_n}{\partial p_k \partial p_l}.
    \label{eq:g_kl}
\end{equation}
Because the residual is linear in the energy, the mixed derivative involving two
wavefunction parameters and the energy is
\begin{equation}
    G_{n,klE}
    =
    \frac{\partial^3 G_n}{\partial p_k \partial p_l \partial E}
    =
    -
    \frac{\partial^2 f_n}{\partial p_k \partial p_l}.
    \label{eq:g_kle}
\end{equation}
Similarly, because the Hamiltonian depends linearly on \(\lambda\), the mixed
derivative involving two wavefunction parameters and \(\lambda\) is
\begin{equation}
    G_{n,kl\lambda}
    =
    \frac{\partial^3 G_n}{\partial p_k \partial p_l \partial \lambda}
    =
    \sum_{m \in S}
    \frac{\partial^2 f_m}{\partial p_k \partial p_l}
    \langle n | \hat{V} | m \rangle .
    \label{eq:g_kllambda}
\end{equation}
Equations~\eqref{eq:g_kl}--\eqref{eq:g_kllambda} are the additional derivative
tensors required by the present higher-order FANPT approximation.

The remaining mixed derivatives needed below are
\begin{equation}
    G_{n,kE}
    =
    \frac{\partial^2 G_n}{\partial p_k \partial E}
    =
    -
    \frac{\partial f_n}{\partial p_k},
    \label{eq:g_ke}
\end{equation}
and
\begin{equation}
    G_{n,k\lambda}
    =
    \frac{\partial^2 G_n}{\partial p_k \partial \lambda}
    =
    \sum_{m \in S}
    \frac{\partial f_m}{\partial p_k}
    \langle n | \hat{V} | m \rangle .
    \label{eq:g_klambda}
\end{equation}

\subsection{Second and third order FANPT response equations}

Differentiating the residual condition
\begin{equation}
    G_n(\lambda,E_\lambda,\mathbf{P}_\lambda)=0
\end{equation}
with respect to \(\lambda\) gives a sequence of linear equations for the response
quantities \(p_k^{(r)}\) and \(E^{(r)}\). At each order, the same first-order
response matrix appears on the left-hand side:
\begin{equation}
    \sum_k G_{n,k} p_k^{(r)}
    +
    G_{n,E} E^{(r)}
    =
    -B_n^{(r)}.
    \label{eq:general_response}
\end{equation}
The order-dependent vector \(B_n^{(r)}\) depends only on lower-order responses and
on derivatives of the residual.

At first order,
\begin{equation}
B_n^{(1)} = G_{n,\lambda},
\end{equation}
so that
\begin{equation}
\sum_k G_{n,k}p_k^{(1)} + G_{n,E}E^{(1)} = -G_{n,\lambda}.
\label{eq:first_order_fanpt}
\end{equation}
At second order, the present approximation gives
\begin{equation}
B_n^{(2)} =
2\sum_k G_{n,k\lambda}p_k^{(1)}
+2\sum_k G_{n,kE}p_k^{(1)}E^{(1)}
+\sum_{k,l}G_{n,kl}p_k^{(1)}p_l^{(1)} .
\label{eq:second_order_fanpt}
\end{equation}
Therefore,
\begin{equation}
\sum_k G_{n,k}p_k^{(2)} + G_{n,E}E^{(2)}
= -B_n^{(2)} .
\end{equation}
The final term in Eq.~\eqref{eq:second_order_fanpt} is absent in the original
quasilinear FANPT approximation and arises from the retained overlap Hessian.
At third order,
\begin{align}
B_n^{(3)} = 3\Bigg[
&\sum_k G_{n,kE}\left(p_k^{(1)}E^{(2)}+p_k^{(2)}E^{(1)}\right)
+\sum_k G_{n,k\lambda}p_k^{(2)} \nonumber \\
&+\sum_{k,l}G_{n,kl}p_k^{(2)}p_l^{(1)}
+\sum_{k,l}G_{n,klE}p_k^{(1)}p_l^{(1)}E^{(1)}
+\sum_{k,l}G_{n,kl\lambda}p_k^{(1)}p_l^{(1)}
\Bigg],
\end{align}
\begin{equation}
\sum_k G_{n,k}p_k^{(3)} + G_{n,E}E^{(3)}
= -B_n^{(3)} .
\end{equation}
Compared with the original approximation, the new contributions are the terms
containing \(G_{n,kl}\), \(G_{n,klE}\), and \(G_{n,kl\lambda}\).

At fourth order,
\begin{align}
B_n^{(4)} =
&\sum_k G_{n,kE}
\left(
4p_k^{(3)}E^{(1)}
+6p_k^{(2)}E^{(2)}
+4p_k^{(1)}E^{(3)}
\right)
+4\sum_k G_{n,k\lambda}p_k^{(3)}
\nonumber \\
&+\sum_{k,l}G_{n,kl}
\left(
4p_k^{(3)}p_l^{(1)}
+3p_k^{(2)}p_l^{(2)}
\right)
+12\sum_{k,l}G_{n,kl\lambda}p_k^{(2)}p_l^{(1)}
\nonumber \\
&+\sum_{k,l}G_{n,klE}
\left(
12p_k^{(2)}p_l^{(1)}E^{(1)}
+6p_k^{(1)}p_l^{(1)}E^{(2)}
\right).
\end{align}
\begin{equation}
\sum_k G_{n,k}p_k^{(4)} + G_{n,E}E^{(4)}
= -B_n^{(4)} 
\label{eq:fourth_order_fanpt}
\end{equation}
Equations~\eqref{eq:first_order_fanpt}--\eqref{eq:fourth_order_fanpt} show that
the present approximation preserves the same left-hand-side response matrix as the
original FANPT formulation. The only modification is the inclusion of additional
terms in the constant vector through the tensors
\(G_{n,kl}\), \(G_{n,klE}\), and \(G_{n,kl\lambda}\).

\subsection{Overlap Hessian for coupled-cluster wavefunctions}

The practical implementation of the present approximation requires the overlap
Hessian
\begin{equation}
    \frac{\partial^2 f_m}{\partial p_k \partial p_l}.
\end{equation}
For linear CI wavefunctions, the determinant overlaps are linear in the CI
coefficients, and therefore
\begin{equation}
    \frac{\partial^2 f_m}{\partial p_k \partial p_l}=0.
\end{equation}
Thus, for CI wavefunctions, the present higher-order approximation is identical to
the original quasilinear FANPT approximation.

For coupled-cluster wavefunctions, the overlaps are nonlinear functions of the
cluster amplitudes. In the product representation, the coupled-cluster wavefunction
can be written as
\begin{equation}
    |\Psi_{\mathrm{CC}}\rangle
    =
    \prod_{\mu}
    \left(1+t_\mu \hat{\tau}_\mu\right)
    |\Phi_{\mathrm{ref}}\rangle,
    \label{eq:cc_product_form}
\end{equation}
where \(\hat{\tau}_\mu\) is an excitation operator and \(t_\mu\) is the
corresponding cluster amplitude. For a determinant \(|m\rangle\), the overlap can
be expressed as a sum over all valid excitation paths \(\pi\) that connect the
reference determinant to \(|m\rangle\):
\begin{equation}
    f_m(\mathbf{t})
    =
    \sum_{\pi \in \mathcal{P}_m}
    s_\pi
    \prod_{\mu \in \pi} t_\mu.
    \label{eq:cc_overlap_paths}
\end{equation}
Here, \(\mathcal{P}_m\) is the set of valid excitation paths leading to determinant
\(|m\rangle\), and \(s_\pi\) is the fermionic sign associated with path \(\pi\).

The first derivative of the overlap with respect to a cluster amplitude \(t_k\) is
\begin{equation}
    \frac{\partial f_m}{\partial t_k}
    =
    \sum_{\pi \in \mathcal{P}_m}
    s_\pi
    \delta_{k\in\pi}
    \prod_{\mu \in \pi,\,\mu\neq k} t_\mu,
    \label{eq:cc_overlap_gradient}
\end{equation}
where \(\delta_{k\in\pi}=1\) if amplitude \(t_k\) appears in path \(\pi\), and zero
otherwise. The second derivative is
\begin{equation}
    \frac{\partial^2 f_m}{\partial t_k \partial t_l}
    =
    \sum_{\pi \in \mathcal{P}_m}
    s_\pi
    \delta_{k\in\pi}
    \delta_{l\in\pi}
    \prod_{\mu \in \pi,\,\mu\neq k,l} t_\mu,
    \qquad k\neq l.
    \label{eq:cc_overlap_hessian}
\end{equation}
For the usual case in which a valid excitation path contains a given excitation
operator at most once, the diagonal Hessian elements vanish,
\begin{equation}
    \frac{\partial^2 f_m}{\partial t_k^2}=0.
\end{equation}
Equation~\eqref{eq:cc_overlap_hessian} is the central new wavefunction quantity
required for the newer approximation.

Once the overlap Hessian is available, the additional FANPT tensors
\(G_{n,kl}\), \(G_{n,klE}\), and \(G_{n,kl\lambda}\) can be constructed using
Eqs.~\eqref{eq:g_kl}--\eqref{eq:g_kllambda}. These tensors modify the
order-dependent constant vector in the FANPT response equations while preserving
the same response matrix used in the original formulation.

\subsection{FANPT methodology}
\begin{figure}[htbp]
    \centering
    \includegraphics[scale=0.2]{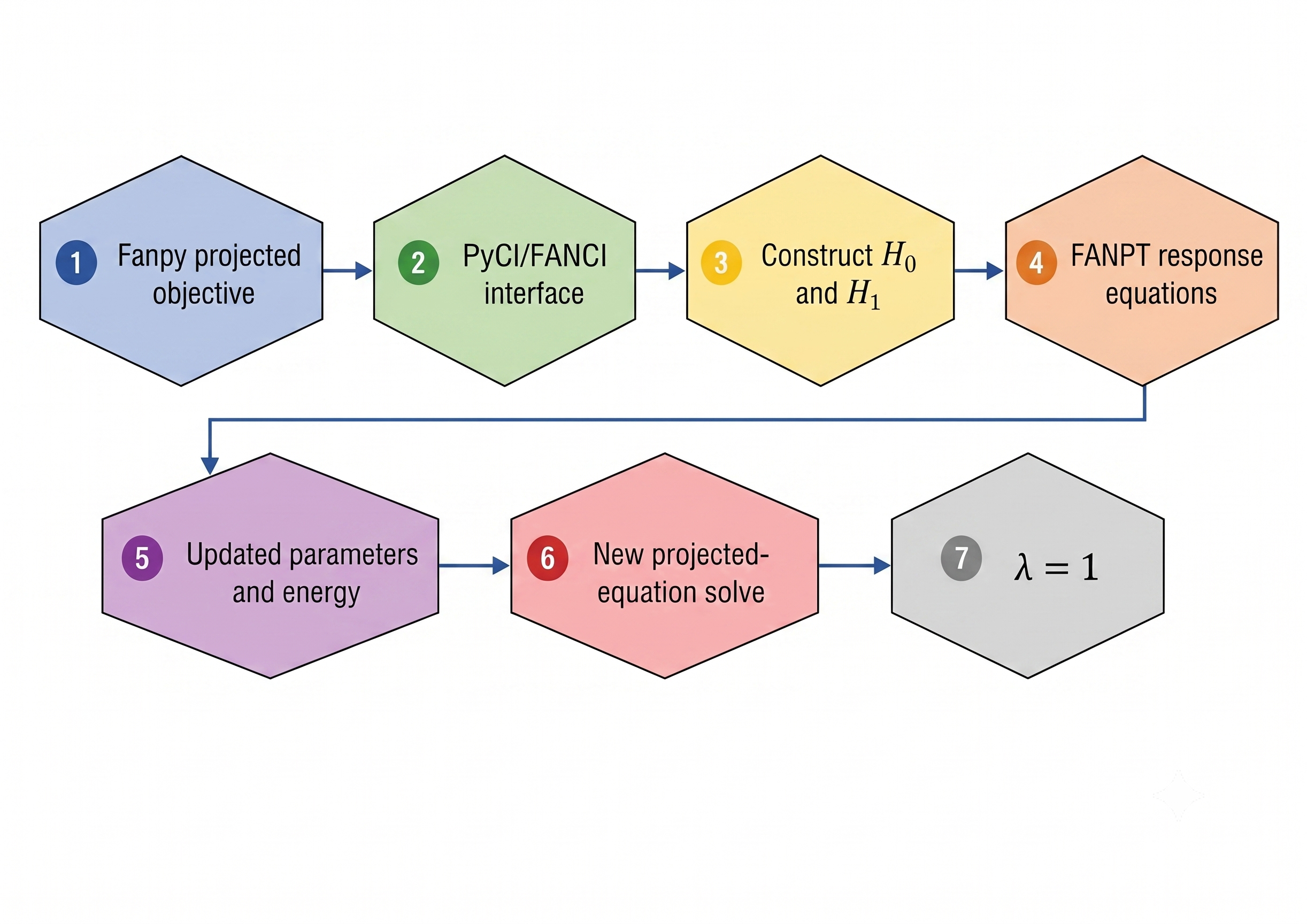}
    \caption{Schematic workflow of the FANPT  implementation in Fanpy.}
    \label{fig:fanpt_workflow}
\end{figure}

FANPT proceeds by evolving the solution from $\lambda = 0$ to $\lambda = 1$
through a sequence of intermediate Hamiltonians. At each value of $\lambda$,
the FANPT response equations are solved to predict the changes in the
wavefunction parameters and, when the energy is treated as an active variable,
the corresponding energy correction. These predicted quantities provide an
initial guess for solving the FANCI projected Schrödinger equations at the next
value of $\lambda$. The updated FANCI solution is then used as the starting
point for the next FANPT step. This procedure is repeated until $\lambda = 1$,
where the Hamiltonian corresponds to the full target molecular Hamiltonian.
In this sense, FANPT serves as a continuation strategy that guides the FANCI
optimization from an easily solved reference problem to the final correlated
electronic-structure problem. The overall workflow schematic is shown in Figure~\ref{fig:fanpt_workflow}. 

\section{Computational Details and Model Systems}

The FANPT implementation was tested on \ce{LiH} at its equilibrium geometry and on the \ce{BeH2} insertion coordinate. All calculations used the STO-6G basis set with molecular orbitals obtained from restricted Hartree--Fock (RHF) calculations. The one- and two-electron molecular integrals were generated using PySCF \cite{cg2020,cg2018,sq2015} and then used in the FanPy \cite{fanpy_github, ap2023,richerPyCIPythonscriptableLibrary2024a, fanpy_2026} implementation of FANCI and FANPT.

For \ce{BeH2}, we used the standard $C_{2v}$ insertion model, where the \ce{Be} atom is placed at the origin and the two hydrogen atoms are placed symmetrically as $(0,\pm y,z)$.\cite{pg1983} This system is a useful test because it contains both weak and strong correlation effects arising from the near-degeneracy of the Be $2s$ and $2p$ orbitals along the insertion path. The geometries used in this work are listed in \autoref{tab:beh2_coordinates}.

\begin{table}[h!]
\centering
\begin{tabular}{c c c c}
\hline
\textbf{Point} & \textbf{$x$} & \textbf{$y$} & \textbf{$z$} \\
\hline
A & 0.0 & $\pm 2.54$  & 0.0  \\
B & 0.0 & $\pm 2.08$  & 1.0  \\
C & 0.0 & $\pm 1.62$  & 2.0  \\
D & 0.0 & $\pm 1.39$  & 2.5  \\
E & 0.0 & $\pm 1.275$ & 2.75 \\
F & 0.0 & $\pm 1.16$  & 3.0  \\
G & 0.0 & $\pm 0.93$  & 3.5  \\
H & 0.0 & $\pm 0.70$  & 4.0  \\
I & 0.0 & $\pm 0.70$  & 6.0  \\
J & 0.0 & $\pm 0.70$  & 20.0 \\
\hline
\end{tabular}
\caption{Cartesian coordinates of the \ce{BeH2} insertion geometries in Bohr. The \ce{Be} atom is placed at the origin, and the hydrogen atoms are placed at $(0,\pm y,z)$.}
\label{tab:beh2_coordinates}
\end{table}

The wavefunctions used here were seniority-restricted coupled-cluster ansatzes.\cite{mqr2024,lra_2026_1} The general coupled-cluster form is
\begin{equation}
    \ket{\Psi_{\mathrm{CC}}} = e^{\hat{T}}\ket{\Phi_0},
\end{equation}
where $\ket{\Phi_0}$ is the RHF reference determinant. In compact notation, the two seniority-restricted operators considered in this work are
\begin{align}
    \hat{T}_{\mathrm{CCSD(0)}} 
    &= \hat{T}_{2}^{(0)}, \\
    \hat{T}_{\mathrm{CCSDT(2)Q(0)}} 
    &= \hat{T}_{1} + \hat{T}_{2} + \hat{T}_{3}^{(2)} + \hat{T}_{4}^{(0)} .
\end{align}
The superscript denotes the maximum seniority included in that excitation manifold. Thus, CCSD(0) provides a compact seniority-zero model, while CCSDT(2)Q(0) includes higher excitation-rank terms and serves as a more nonlinear test of the higher-order FANPT approximation.

All FANPT calculations were performed using a development version of FanPy. For each system, independently optimized FANCI wavefunctions were first obtained at the relevant geometries. FANPT was then used to propagate the wavefunction parameters along the interpolation parameter $\lambda$, from the reference Hamiltonian at $\lambda=0$ to the target Hamiltonian at $\lambda=1$. At each intermediate step, the FANPT response equations were solved to generate corrections to the wavefunction parameters and, when active, to the energy.

The second- and third-order FANPT expansions were compared using the original quasilinear approximation and the second-derivative-corrected approximation. Performance was assessed by comparing the FANPT-propagated guess with the independently optimized FANCI solution at the same value of $\lambda$. The main diagnostics were the same-$\lambda$ Frobenius norm of the parameter difference, the maximum parameter deviation, and the energy difference.

\section{Results and Discussion}
\subsection{LiH}
As an initial test of the second order corrected FANPT approximation, we considered LiH at its equilibrium bond length, $r=1.6081$~\AA, using the CCSD(0) wavefunction with 100 equally spaced steps along the adiabatic connection. The original quasilinear approximation order, denoted here as qao = 2, and the second derivative corrected approximation, qao = 3, gave nearly identical final energies. The final total energies were $-7.9671658141$ and $-7.9671656982$~hartree for qao = 2 and qao = 3, respectively, differing by only $1.2 \times 10^{-7}$~hartree. Thus, for this weakly correlated equilibrium geometry, the inclusion of the overlap Hessian does not significantly change the final converged energy.

However, the intermediate continuation behavior is improved substantially by the second-derivative correction. To quantify this behavior, we compare the FANPT-propagated parameter vector at each intermediate value of $\lambda$ with the corresponding parameter vector obtained after fully solving the FANCI projected Schrödinger equations at the same value of $\lambda$. The same-$\lambda$ Frobenius norm therefore measures the size of the correction required to relax the FANPT guess to the optimized FANCI solution at that point. Smaller same-$\lambda$ norms indicate that the FANPT propagation provides a more accurate initial guess and that the continuation path follows the optimized FANCI solution more closely. 

The mean same-$\lambda$ Frobenius norm of the parameter correction decreases from $1.26 \times 10^{-1}$ for qao = 2 to $1.72 \times 10^{-2}$ for qao = 3, while the maximum same-$\lambda$ parameter deviation decreases from $8.11 \times 10^{-1}$ to $6.57 \times 10^{-2}$. Selected points along the adiabatic connection where the qao = 2 trajectory exhibits large parameter-space excursions are shown in Table~\ref{tab:lih_qao_comparison}. At these values of $\lambda$, the qao = 3 correction is consistently much smaller, indicating that the perturbative prediction lies closer to the optimized same-$\lambda$ solution. For example, at $\lambda=0.32$, the same-$\lambda$ parameter norm is $8.11 \times 10^{-1}$ for qao = 2, but only $1.09 \times 10^{-2}$ for qao = 3. This suggests that the quasilinear approximation can occasionally produce updates that are far from the local optimized solution, even though the nonlinear solver eventually corrects the trajectory. In contrast, retaining the leading overlap-Hessian contribution gives a smoother continuation path. This improved stability is also reflected in the optimization effort: excluding the initial $\lambda=0$ solve, qao = 3 required 1299 total function evaluations compared with 1378 for qao = 2, and the wall time decreased from 118.2 to 102.4 seconds. These results suggest that, even when the final energy is relatively insensitive to the additional nonlinear terms, the second derivative corrected FANPT approximation can improve the robustness and efficiency of the continuation procedure for nonlinear wavefunction ansätze.

\begin{table}[h]
\centering
\caption{Selected same-$\lambda$ parameter deviations for the LiH equilibrium calculation using 100 FANPT steps. The listed points correspond to large parameter-space excursions in the qao = 2 trajectory.}
\label{tab:lih_qao_comparison}
\begin{tabular}{cccc}
\hline
$\lambda$ & 
qao = 2 same-$\lambda$ norm & 
qao = 3 same-$\lambda$ norm & 
Reduction factor \\
\hline
0.03 & $7.45 \times 10^{-1}$ & $4.29 \times 10^{-3}$ & 174 \\
0.32 & $8.11 \times 10^{-1}$ & $1.09 \times 10^{-2}$ & 75 \\
0.42 & $5.44 \times 10^{-1}$ & $6.06 \times 10^{-3}$ & 90 \\
0.60 & $6.20 \times 10^{-1}$ & $2.97 \times 10^{-2}$ & 21 \\
0.63 & $7.14 \times 10^{-1}$ & $5.67 \times 10^{-2}$ & 13 \\
\hline
\end{tabular}
\end{table}
For the third-order FANPT expansion with 100 steps, the same qualitative trend is observed, although the improvement from qao = 3 is less dramatic than in the second-order case. The final total energies from qao = 2 and qao = 3 are nearly identical, $-7.9671658905$ and $-7.9671657052$~hartree, respectively, differing by only $1.85 \times 10^{-7}$~hartree. Thus, as in the second-order calculation, the final energy is not strongly affected by the inclusion of the overlap-Hessian terms. However, the qao = 3 trajectory is smoother in parameter space. The mean same-$\lambda$ Frobenius norm of the parameter correction decreases from $1.67 \times 10^{-2}$ for qao = 2 to $1.01 \times 10^{-2}$ for qao = 3, and the maximum same-$\lambda$ parameter deviation decreases from $1.11 \times 10^{-1}$ to $3.95 \times 10^{-2}$. Representative points where qao = 2 gives noticeably larger same-$\lambda$ corrections are shown in Table~\ref{tab:lih_order3_qao_comparison}. The largest reduction occurs at $\lambda=0.48$, where the same-$\lambda$ norm decreases from $6.25 \times 10^{-2}$ for qao = 2 to $2.77 \times 10^{-3}$ for qao = 3. The optimization effort is also reduced, with qao = 3 requiring 1226 total function evaluations compared with 1330 for qao = 2. These results indicate that, for the third-order expansion, the second derivative corrected approximation again provides a more stable continuation path, even though both approaches converge to essentially the same final energy.
\begin{table}[h]
\centering
\caption{Selected same-$\lambda$ parameter deviations for the LiH equilibrium calculation using a third-order FANPT expansion with 100 steps. The listed points correspond to values of $\lambda$ where qao = 2 shows noticeably larger parameter corrections than qao = 3.}
\label{tab:lih_order3_qao_comparison}
\begin{tabular}{cccc}
\hline
$\lambda$ &
qao = 2 same-$\lambda$ norm &
qao = 3 same-$\lambda$ norm &
Reduction factor \\
\hline
0.48 & $6.25 \times 10^{-2}$ & $2.77 \times 10^{-3}$ & 22.5 \\
0.76 & $1.11 \times 10^{-1}$ & $8.40 \times 10^{-3}$ & 13.2 \\
0.69 & $6.96 \times 10^{-2}$ & $1.12 \times 10^{-2}$ & 6.2 \\
0.79 & $8.43 \times 10^{-2}$ & $1.70 \times 10^{-2}$ & 5.0 \\
0.52 & $5.10 \times 10^{-2}$ & $1.38 \times 10^{-2}$ & 3.7 \\
\hline
\end{tabular}
\end{table}

\subsection{BeH$_2$}

We next considered the BeH$_2$ insertion coordinate as a more challenging nonlinear test. 
Unlike the LiH equilibrium calculations discussed above, this system contains six electrons and samples a broader range of correlation regimes along the reaction coordinate. 
We also reduced the number of FANPT continuation steps, so each perturbative update spans a larger interval in $\lambda$. 
This makes BeH$_2$ a useful test for comparing the quasilinear qao $=2$ approximation with the second derivative corrected qao $=3$ approximation.

\begin{figure}[h]
    \centering
    \begin{subfigure}[b]{0.45\textwidth}
        \centering
        \includegraphics[scale=0.48]{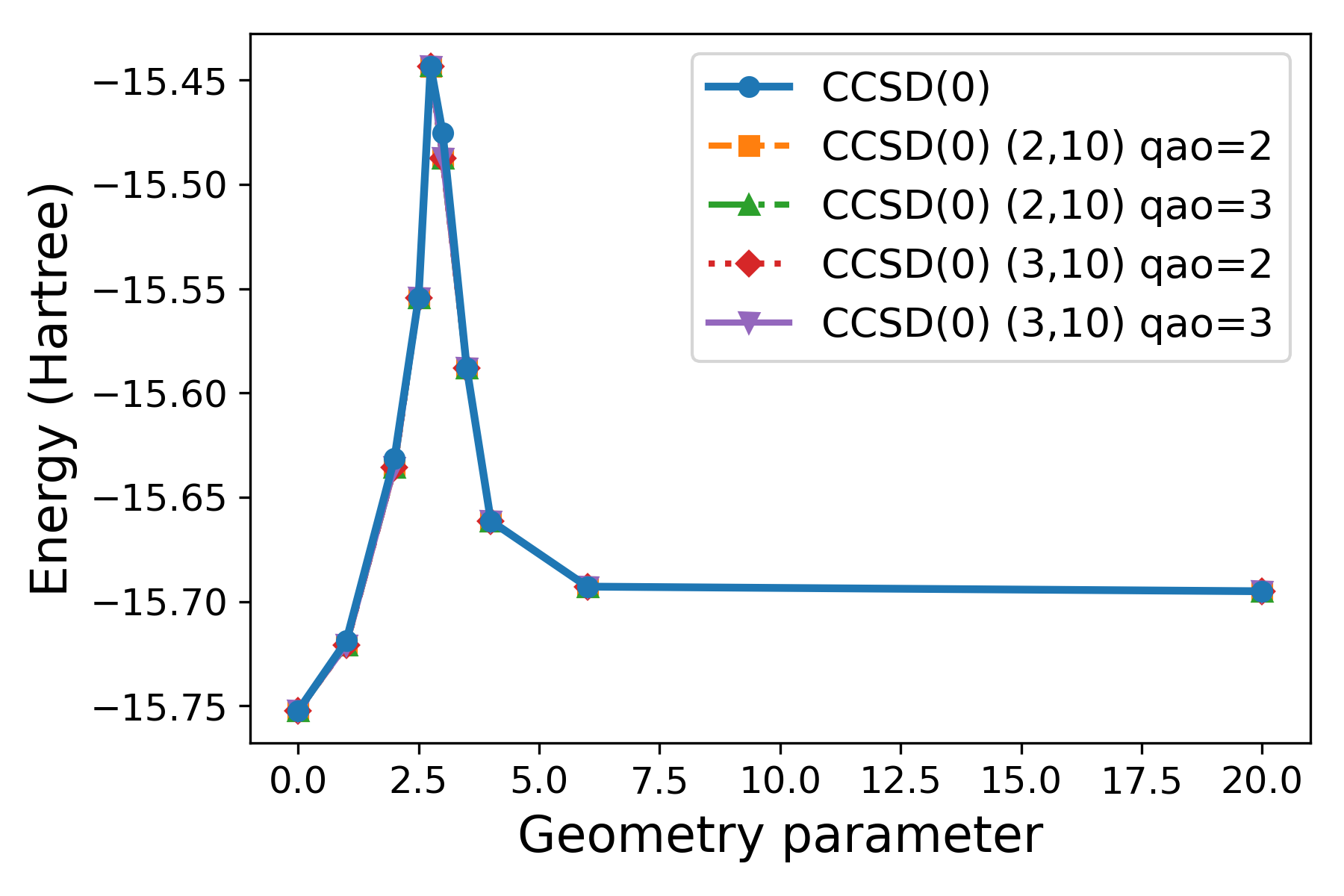}
        \caption{}
        \label{fig:beh2_ccsdsen0_energy}
    \end{subfigure}
    \hfill
    \begin{subfigure}[b]{0.45\textwidth}
        \centering
        \includegraphics[scale=0.48]{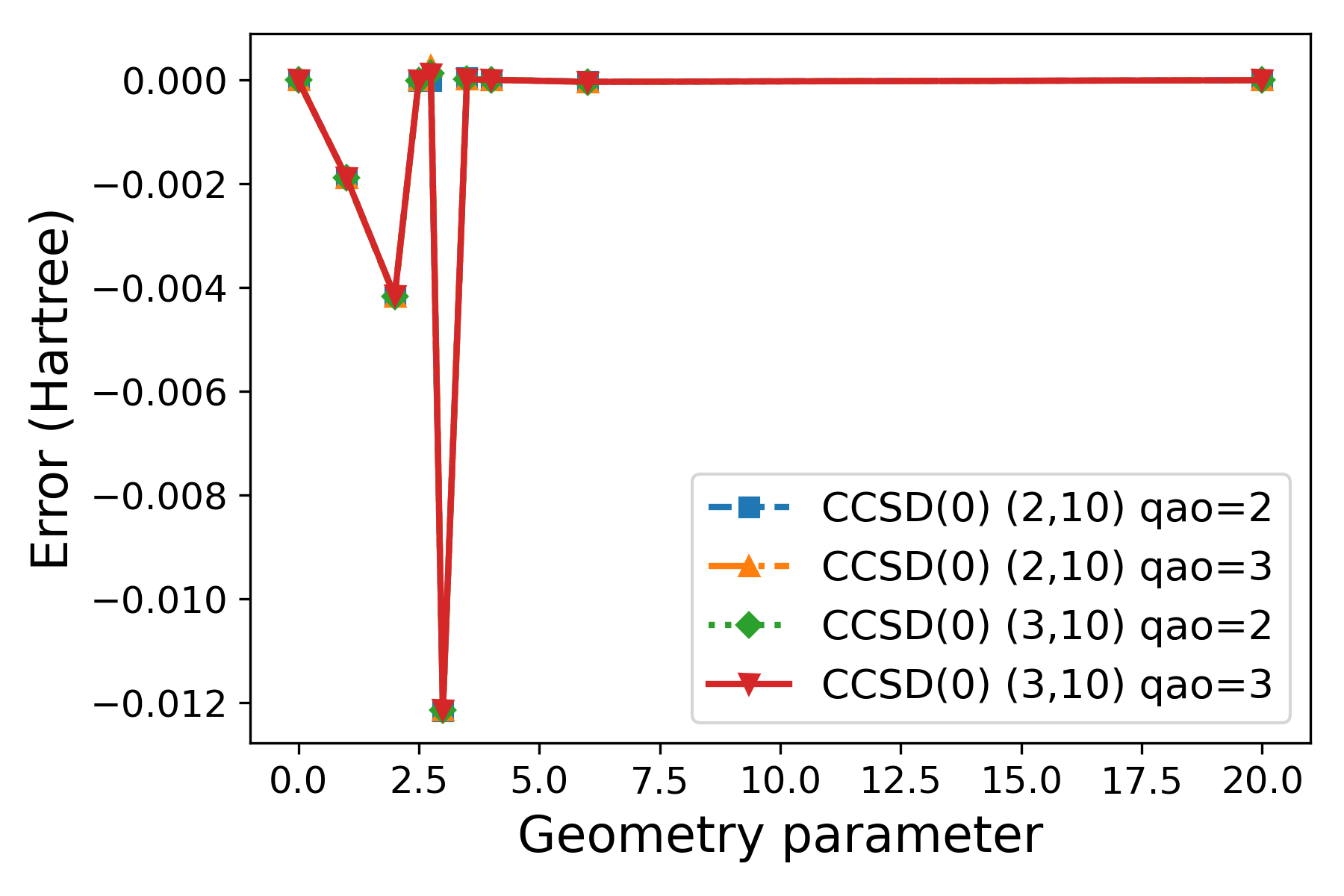}
        \caption{}
        \label{fig:beh2_ccsdsen0_error}
    \end{subfigure}
    \caption{FANPT results for the $C_{2v}$ insertion of \ce{Be} into \ce{H2} using the CCSD(0) wavefunction. 
    (a) Total energies and (b) energy differences relative to the independently optimized CCSD(0) wavefunction energies. 
    All calculations used the STO-6G basis set.}
    \label{fig:results-BeH2-CCSD(0)}
\end{figure}

We first examine the CCSD(0) wavefunction. 
Before discussing the same-$\lambda$ norm diagnostics, we compare the FANPT energies with the independently optimized CCSD(0) wavefunction energies along the BeH$_2$ coordinate, as shown in Fig.~\ref{fig:results-BeH2-CCSD(0)}. 
For all reduced-step FANPT calculations considered here, the final FANPT energies remain close to the optimized CCSD(0) energies. 
The deviations are typically on the order of $10^{-3}$ hartree, with the largest errors reaching approximately $10^{-2}$ hartree near the more difficult region of the insertion coordinate. 
The energy errors are also very similar for qao $=2$ and qao $=3$.

The same-$\lambda$ diagnostics compare the FANPT-predicted solution at each value of $\lambda$ with the independently optimized FANCI solution at the same $\lambda$. 
As shown in Table~\ref{tab:beh2_CCSD(0)_overall_stats}, qao $=3$ substantially improves these diagnostics for the second-order reduced-step CCSD(0) calculation. 
The mean Frobenius norm decreases from 4.32 for qao $=2$ to 0.567 for qao $=3$, while the maximum norm decreases from 99.88 to 5.24. 
The mean energy change is also reduced from $6.40 \times 10^{-3}$ to $2.88 \times 10^{-3}$ hartree.

\begin{table}[h!]
\centering
\caption{Overall same-$\lambda$ diagnostics for the second-order BeH$_2$/CCSD(0) calculation using a reduced number of FANPT continuation steps.}
\label{tab:beh2_CCSD(0)_overall_stats}
\begin{tabular}{lccc}
\hline
\textbf{Diagnostic} & \textbf{qao $=2$} & \textbf{qao $=3$} & \textbf{Improvement} \\
\hline
Mean Frobenius norm        & 4.32 & 0.567 & $7.6\times$ \\
Median Frobenius norm      & 0.257 & 0.119 & $2.2\times$ \\
Maximum Frobenius norm     & 99.88 & 5.24 & $19.1\times$ \\
Mean parameter norm        & 4.32 & 0.567 & $7.6\times$ \\
Maximum parameter norm     & 99.88 & 5.24 & $19.1\times$ \\
Mean energy change         & $6.40 \times 10^{-3}$ & $2.88 \times 10^{-3}$ & $2.2\times$ \\
Maximum energy change      & $3.23 \times 10^{-1}$ & $1.30 \times 10^{-2}$ & $24.8\times$ \\
\hline
\end{tabular}
\end{table}

The largest differences occur near the most difficult region of the BeH$_2$ coordinate, especially around $r=3.0$ and $r=3.5$. 
Representative points are listed in Table~\ref{tab:beh2_CCSD(0)_selected_norms}. 
At $r=3.5$ and $\lambda=1.0$, the same-$\lambda$ Frobenius norm decreases from 99.88 for qao $=2$ to $7.09 \times 10^{-2}$ for qao $=3$. 
At $r=3.0$ and $\lambda=0.3$, the norm decreases from 38.42 to $1.42 \times 10^{-1}$. 
This point also gives the largest same-$\lambda$ energy change for qao $=2$: $3.23 \times 10^{-1}$ hartree, compared with only $5.74 \times 10^{-4}$ hartree for qao $=3$.

\begin{table}[h!]
\centering
\caption{Selected same-$\lambda$ Frobenius norms for the second-order BeH$_2$/CCSD(0) reduced-step calculation.}
\label{tab:beh2_CCSD(0)_selected_norms}
\begin{tabular}{ccccc}
\hline
\textbf{$r$} & \textbf{$\lambda$} & \textbf{qao $=2$ norm} & \textbf{qao $=3$ norm} & \textbf{Reduction factor} \\
\hline
3.5 & 1.0 & $9.99 \times 10^{1}$ & $7.09 \times 10^{-2}$ & $1.41 \times 10^{3}$ \\
3.5 & 0.7 & $2.26 \times 10^{1}$ & $1.80 \times 10^{-2}$ & $1.26 \times 10^{3}$ \\
3.5 & 0.9 & $6.71 \times 10^{1}$ & $7.14 \times 10^{-2}$ & $9.40 \times 10^{2}$ \\
3.5 & 0.8 & $4.51 \times 10^{1}$ & $5.07 \times 10^{-2}$ & $8.89 \times 10^{2}$ \\
3.5 & 0.3 & $2.64 \times 10^{1}$ & $4.91 \times 10^{-2}$ & $5.38 \times 10^{2}$ \\
3.0 & 0.3 & $3.84 \times 10^{1}$ & $1.42 \times 10^{-1}$ & $2.71 \times 10^{2}$ \\
\hline
\end{tabular}
\end{table}

These reductions show that qao $=2$ can produce very large parameter updates when the continuation path passes through difficult regions of the BeH$_2$ coordinate. 
The qao $=3$ correction suppresses these large excursions and keeps the FANPT prediction closer to the optimized same-$\lambda$ solution. 
This trend is also reflected in the number of large-norm events: qao $=2$ gives 29 points with norms larger than 1, 14 larger than 5, and 11 larger than 10. 
For qao $=3$, these counts decrease to 14, 1, and 0, respectively.

The improvement is not uniform at every geometry. 
For example, near $r=2.5$, qao $=3$ gives larger same-$\lambda$ norms than qao $=2$, with its maximum norm of 5.24 occurring at $r=2.5$ and $\lambda=0.7$. 
Smaller deteriorations are also observed near $r=1.0$ and partially near $r=6.0$. 
However, these deviations remain modest compared with the severe qao $=2$ excursions near $r=3.0$ and $r=3.5$. 
Thus, although qao $=2$ can be locally better at some points, qao $=3$ gives the more reliable global continuation behavior.

We also tested BeH$_2$/CCSD(0) using a third-order FANPT expansion with only 10 continuation steps. 
This is a more demanding calculation because the perturbative updates are taken over larger intervals in $\lambda$. 
As summarized in Table~\ref{tab:beh2_CCSD(0)_order3_steps10_overall}, qao $=3$ again gives better global stability. 
The mean Frobenius norm decreases from 6.516 to 0.377, and the maximum norm decreases from 174.55 to 5.44. 
The mean energy changes are very similar for the two approximations, which again shows that the main difference is in the parameter-space trajectory rather than in the final energy.

\begin{table}[h!]
\centering
\caption{Overall same-$\lambda$ diagnostics for the third-order BeH$_2$/CCSD(0) calculation using 10 FANPT continuation steps.}
\label{tab:beh2_CCSD(0)_order3_steps10_overall}
\begin{tabular}{lccc}
\hline
\textbf{Diagnostic} & \textbf{qao $=2$} & \textbf{qao $=3$} & \textbf{Better} \\
\hline
Mean Frobenius norm & 6.516 & 0.377 & qao $=3$ \\
Median Frobenius norm & 0.0787 & 0.0813 & qao $=2$ \\
Maximum Frobenius norm & 174.55 & 5.44 & qao $=3$ \\
Mean parameter norm & 6.516 & 0.377 & qao $=3$ \\
Maximum parameter norm & 174.55 & 5.44 & qao $=3$ \\
Mean energy change & $2.97 \times 10^{-3}$ & $2.88 \times 10^{-3}$ & qao $=3$ \\
Maximum energy change & $1.30 \times 10^{-2}$ & $1.30 \times 10^{-2}$ & Comparable \\
\hline
\end{tabular}
\end{table}

The main advantage of qao $=3$ in the third-order CCSD(0) calculation is the removal of the large qao $=2$ instability near $r=4.0$. 
As shown in Table~\ref{tab:beh2_CCSD(0)_order3_steps10_selected}, the same-$\lambda$ Frobenius norm at $r=4.0$ and $\lambda=1.0$ decreases from 174.55 for qao $=2$ to 0.225 for qao $=3$. 
Similar reductions occur throughout the later part of the adiabatic connection.

\begin{table}[h!]
\centering
\caption{Selected same-$\lambda$ Frobenius norms for the third-order BeH$_2$/CCSD(0) calculation using 10 FANPT continuation steps.}
\label{tab:beh2_CCSD(0)_order3_steps10_selected}
\begin{tabular}{ccccc}
\hline
\textbf{$r$} & \textbf{$\lambda$} & \textbf{qao $=2$ norm} & \textbf{qao $=3$ norm} & \textbf{Reduction factor} \\
\hline
4.0 & 0.9 & $1.27 \times 10^{2}$ & $7.14 \times 10^{-2}$ & $1.78 \times 10^{3}$ \\
4.0 & 0.8 & $9.25 \times 10^{1}$ & $7.13 \times 10^{-2}$ & $1.30 \times 10^{3}$ \\
4.0 & 0.7 & $8.24 \times 10^{1}$ & $7.13 \times 10^{-2}$ & $1.16 \times 10^{3}$ \\
4.0 & 0.6 & $5.51 \times 10^{1}$ & $6.76 \times 10^{-2}$ & $8.15 \times 10^{2}$ \\
4.0 & 1.0 & $1.75 \times 10^{2}$ & $2.25 \times 10^{-1}$ & $7.74 \times 10^{2}$ \\
4.0 & 0.5 & $3.70 \times 10^{1}$ & $7.12 \times 10^{-2}$ & $5.19 \times 10^{2}$ \\
\hline
\end{tabular}
\end{table}

The number of large-norm events supports the same conclusion. 
For qao $=2$, 15 points have same-$\lambda$ Frobenius norms larger than 1, 8 are larger than 5, and 8 are larger than 10. 
For qao $=3$, these counts decrease to 7, 1, and 0, respectively, as shown in Table~\ref{tab:beh2_CCSD(0)_order3_steps10_large_norm_counts}. 
Therefore, for both second- and third-order BeH$_2$/CCSD(0) calculations, qao $=3$ provides the more robust continuation strategy.

\begin{table}[h!]
\centering
\caption{Number of large same-$\lambda$ Frobenius-norm events for the third-order BeH$_2$/CCSD(0) calculation with 10 FANPT continuation steps.}
\label{tab:beh2_CCSD(0)_order3_steps10_large_norm_counts}
\begin{tabular}{ccc}
\hline
\textbf{Threshold} & \textbf{qao $=2$ count} & \textbf{qao $=3$ count} \\
\hline
Norm $> 1$ & 15 & 7 \\
Norm $> 5$ & 8 & 1 \\
Norm $> 10$ & 8 & 0 \\
Norm $> 50$ & 5 & 0 \\
Norm $> 100$ & 2 & 0 \\
\hline
\end{tabular}
\end{table}

\begin{figure}[h]
    \centering
    \begin{subfigure}[b]{0.45\textwidth}
        \centering
        \includegraphics[scale=0.48]{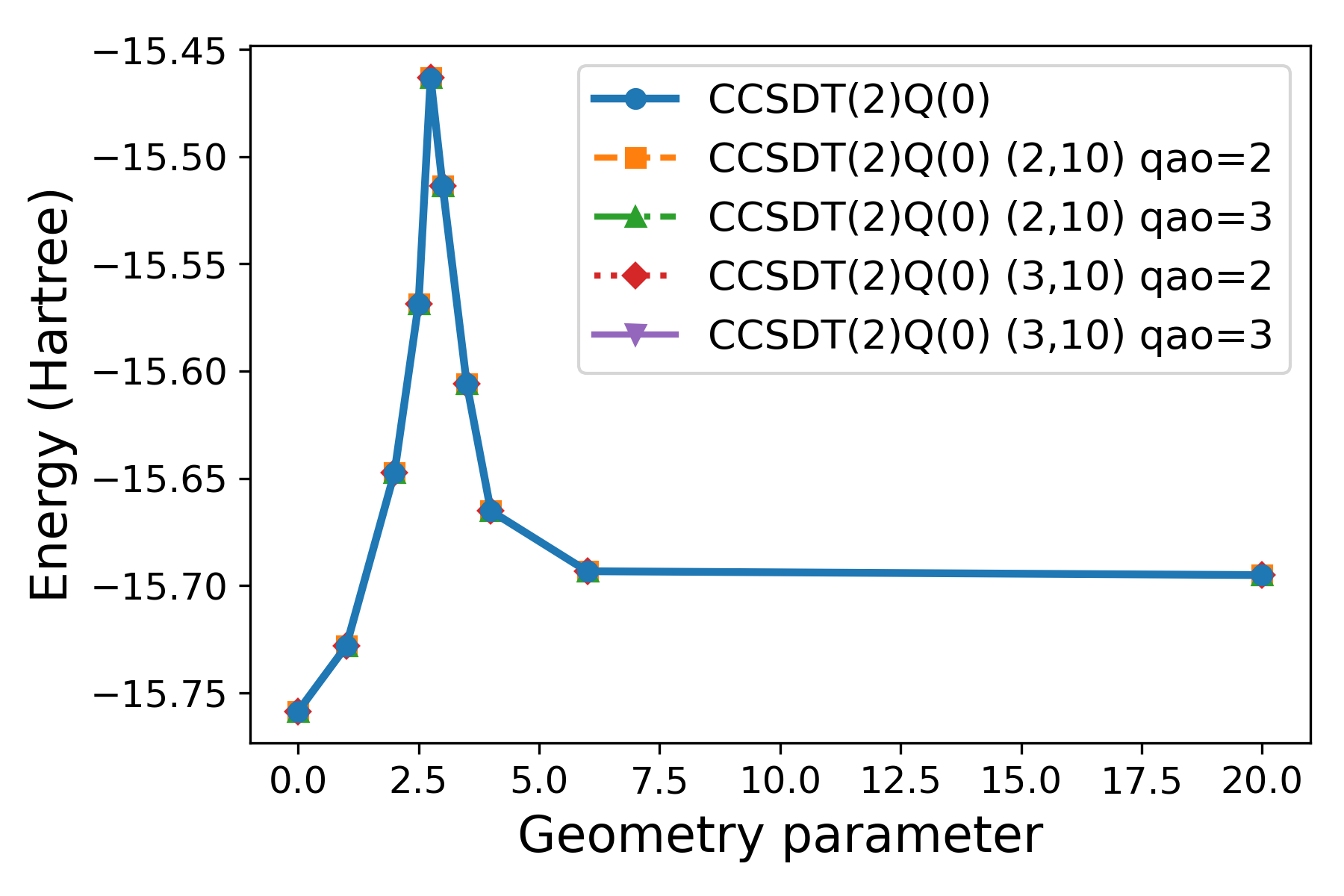}
        \caption{}
        \label{fig:beh2_ccsdtsen2qsen0_energy}
    \end{subfigure}
    \hfill
    \begin{subfigure}[b]{0.45\textwidth}
        \centering
        \includegraphics[scale=0.48]{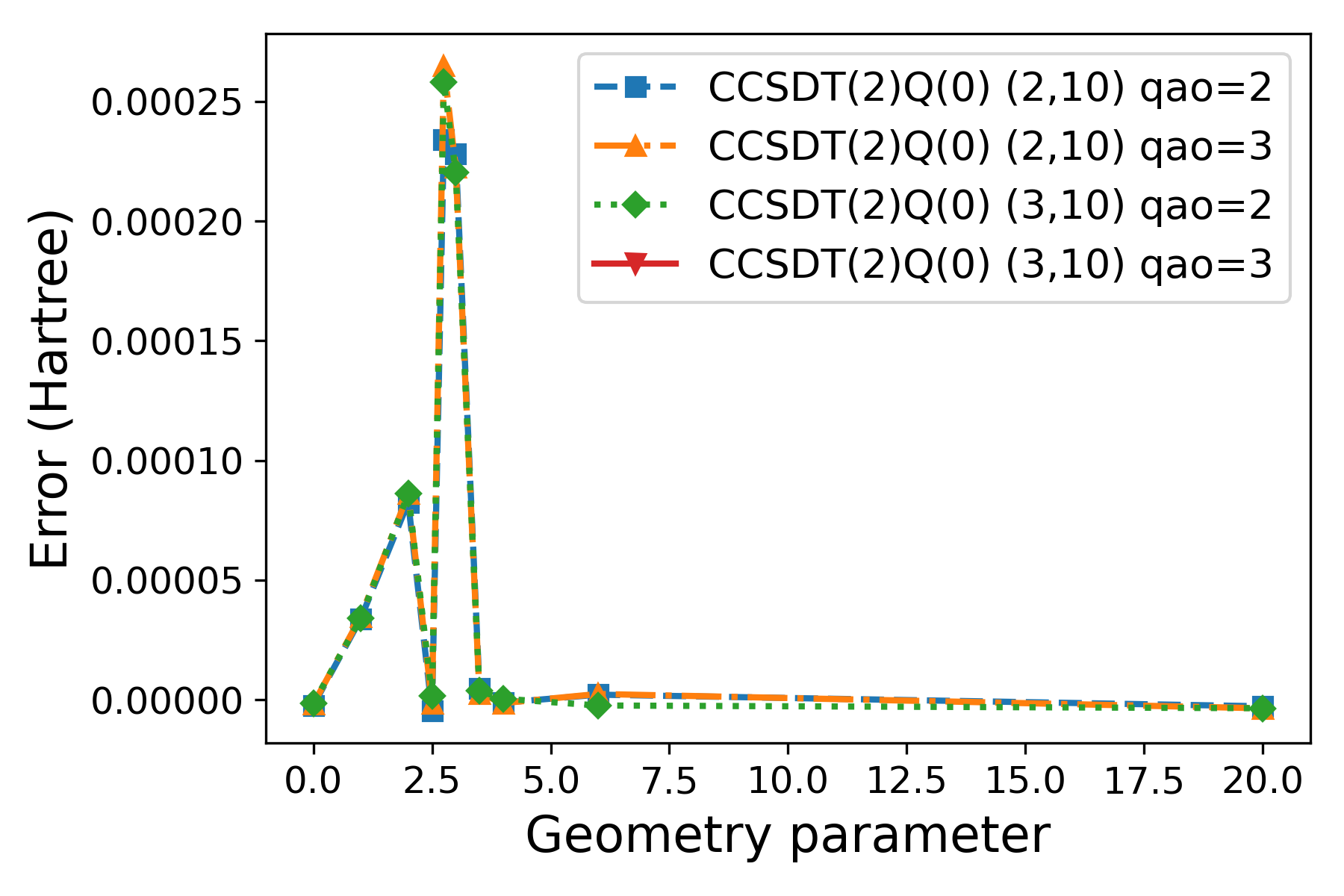}
        \caption{}
        \label{fig:beh2_ccsdtsen2qsen0_error}
    \end{subfigure}
    \caption{FANPT results for the $C_{2v}$ insertion of \ce{Be} into \ce{H2} using the CCSDT(2)Q(0) wavefunction. 
    (a) Total energies and (b) energy differences relative to the independently optimized CCSDT(2)Q(0) wavefunction energies. 
    All calculations used the STO-6G basis set.}
    \label{fig:results-BeH2-CCSDT(2)Q(0)}
\end{figure}

We then repeated the BeH$_2$ test using the more flexible CCSDT(2)Q(0) wavefunction. 
This ansatz includes higher excitation-rank amplitudes than CCSD(0), making it a more demanding test of the FANPT continuation procedure. 
The calculations were again performed with 10 continuation steps. 
As shown in Fig.~\ref{fig:results-BeH2-CCSDT(2)Q(0)}, the final FANPT energies remain close to the independently optimized CCSDT(2)Q(0) energies. 
The energy-error trends again do not strongly distinguish qao $=2$ from qao $=3$, so we focus on the same-$\lambda$ diagnostics.

For the second-order CCSDT(2)Q(0) calculation, the same-$\lambda$ energy changes are nearly identical for qao $=2$ and qao $=3$. 
The mean energy change is $3.51 \times 10^{-3}$ hartree for qao $=2$ and $3.45 \times 10^{-3}$ hartree for qao $=3$, while the maximum values are $1.62 \times 10^{-2}$ and $1.63 \times 10^{-2}$ hartree, respectively. 
In contrast, the parameter-space diagnostics show a clear improvement with qao $=3$. 
As shown in Table~\ref{tab:beh2_CCSDT(2)Q(0)_order2_overall}, the mean Frobenius norm decreases from 10.41 to 0.696, and the maximum norm decreases from 384.56 to 18.53.

\begin{table}[htbp]
\centering
\caption{Overall same-$\lambda$ diagnostics for the second-order BeH$_2$/CCSDT(2)Q(0) calculation using 10 FANPT continuation steps.}
\label{tab:beh2_CCSDT(2)Q(0)_order2_overall}
\begin{tabularx}{0.95\textwidth}{lccc}
\hline
\textbf{Diagnostic} & \textbf{qao $=2$} & \textbf{qao $=3$} & \textbf{Better} \\
\hline
Mean Frobenius norm & 10.41 & 0.696 & qao $=3$ \\
Median Frobenius norm & 0.258 & 0.0713 & qao $=3$ \\
Maximum Frobenius norm & 384.56 & 18.53 & qao $=3$ \\
Mean parameter norm & 10.41 & 0.696 & qao $=3$ \\
Maximum parameter norm & 384.56 & 18.53 & qao $=3$ \\
Mean energy change & $3.51\times 10^{-3}$ & $3.45\times 10^{-3}$ & qao $=3$ \\
Maximum energy change & $1.62\times 10^{-2}$ & $1.63\times 10^{-2}$ & Comparable \\
\hline
\end{tabularx}
\end{table}

The largest reductions occur at stretched geometries, as shown in Table~\ref{tab:beh2_CCSDT(2)Q(0)_order2_selected}. 
At $r=20.0$ and $\lambda=0.9$, the same-$\lambda$ Frobenius norm decreases from 384.56 for qao $=2$ to 1.49 for qao $=3$. 
At $r=4.0$ and $\lambda=1.0$, it decreases from 31.78 to 0.268. 
These results show that qao $=3$ strongly suppresses the largest parameter-space excursions observed with qao $=2$.

\begin{table}[htbp]
\centering
\caption{Selected same-$\lambda$ Frobenius norms for the second-order BeH$_2$/CCSDT(2)Q(0) calculation using 10 FANPT continuation steps.}
\label{tab:beh2_CCSDT(2)Q(0)_order2_selected}
\begin{tabularx}{0.85\textwidth}{ccccc}
\hline
\textbf{$r$} & \textbf{$\lambda$} & \textbf{qao $=2$} & \textbf{qao $=3$} & \textbf{Reduction factor} \\
\hline
20.0 & 0.9 & $3.85\times 10^{2}$ & 1.49 & $2.58\times 10^{2}$ \\
20.0 & 1.0 & $1.50\times 10^{2}$ & 5.96 & $2.51\times 10^{1}$ \\
4.0  & 1.0 & $3.18\times 10^{1}$ & 0.268 & $1.19\times 10^{2}$ \\
6.0  & 1.0 & $2.71\times 10^{1}$ & 18.53 & 1.46 \\
\hline
\end{tabularx}
\end{table}

The large-norm counts lead to the same conclusion. 
For qao $=2$, 33 points have same-$\lambda$ Frobenius norms larger than 1, 21 are larger than 5, and 15 are larger than 10. 
For qao $=3$, these counts decrease to 13, 3, and 2, respectively. 
No qao $=3$ point exceeds a norm of 50, whereas qao $=2$ has three such events. 
Thus, qao $=3$ provides the more reliable continuation path for the second-order CCSDT(2)Q(0) calculation.

Finally, we considered the third-order FANPT expansion for BeH$_2$/CCSDT(2)Q(0), again using 10 continuation steps. 
The energy behavior remains nearly unchanged between qao $=2$ and qao $=3$. 
The mean same-$\lambda$ energy change is $3.50 \times 10^{-3}$ hartree for qao $=2$ and $3.45 \times 10^{-3}$ hartree for qao $=3$, while the maximum energy change is $1.62 \times 10^{-2}$ hartree for both. 
The parameter-space diagnostics, however, again favor qao $=3$. 
As summarized in Table~\ref{tab:beh2_CCSDT(2)Q(0)_order3_overall}, the mean Frobenius norm decreases from 16.40 to 0.665, and the maximum norm decreases from 286.95 to 18.67.

\begin{table}[htbp]
\centering
\caption{Overall same-$\lambda$ diagnostics for the third-order BeH$_2$/CCSDT(2)Q(0) calculation using 10 FANPT continuation steps.}
\label{tab:beh2_CCSDT(2)Q(0)_order3_overall}
\begin{tabularx}{0.95\textwidth}{lccc}
\hline
\textbf{Diagnostic} & \textbf{qao $=2$} & \textbf{qao $=3$} & \textbf{Better} \\
\hline
Mean Frobenius norm & 16.40 & 0.665 & qao $=3$ \\
Median Frobenius norm & 0.396 & 0.125 & qao $=3$ \\
Maximum Frobenius norm & 286.95 & 18.67 & qao $=3$ \\
Mean parameter norm & 16.40 & 0.665 & qao $=3$ \\
Maximum parameter norm & 286.95 & 18.67 & qao $=3$ \\
Mean energy change & $3.50\times 10^{-3}$ & $3.45\times 10^{-3}$ & qao $=3$ \\
Maximum energy change & $1.62\times 10^{-2}$ & $1.62\times 10^{-2}$ & Comparable \\
\hline
\end{tabularx}
\end{table}

The largest improvements again occur at stretched geometries. 
At $r=20.0$ and $\lambda=0.8$, the same-$\lambda$ Frobenius norm decreases from 286.95 for qao $=2$ to 18.67 for qao $=3$. 
At $r=20.0$ and $\lambda=1.0$, it decreases from 239.31 to 1.57, and at $r=4.0$ and $\lambda=1.0$, it decreases from 83.76 to 0.296. 
Representative points are listed in Table~\ref{tab:beh2_CCSDT(2)Q(0)_order3_selected}.

\begin{table}[htbp]
\centering
\caption{Selected same-$\lambda$ Frobenius norms for the third-order BeH$_2$/CCSDT(2)Q(0) calculation using 10 FANPT continuation steps.}
\label{tab:beh2_CCSDT(2)Q(0)_order3_selected}
\begin{tabularx}{0.85\textwidth}{ccccc}
\hline
\textbf{$r$} & \textbf{$\lambda$} & \textbf{qao $=2$} & \textbf{qao $=3$} & \textbf{Reduction factor} \\
\hline
20.0 & 0.8 & $2.87\times 10^{2}$ & 18.67 & 15.4 \\
20.0 & 0.7 & $2.03\times 10^{2}$ & 0.611 & $3.32\times 10^{2}$ \\
20.0 & 1.0 & $2.39\times 10^{2}$ & 1.57 & $1.53\times 10^{2}$ \\
4.0  & 1.0 & $8.38\times 10^{1}$ & 0.296 & $2.83\times 10^{2}$ \\
4.0  & 0.9 & $5.47\times 10^{1}$ & 0.303 & $1.81\times 10^{2}$ \\
\hline
\end{tabularx}
\end{table}

The number of large-norm events further supports this trend. 
For qao $=2$, 40 points have same-$\lambda$ Frobenius norms larger than 1, 21 are larger than 5, and 21 are larger than 10. 
For qao $=3$, these counts decrease to 17, 1, and 1, respectively. 
Therefore, for both second- and third-order FANPT with the CCSDT(2)Q(0) wavefunction, qao $=3$ gives the more stable global continuation path.

Overall, the BeH$_2$ calculations show that the main benefit of qao $=3$ is not a large improvement in the final FANPT energy. 
Instead, qao $=3$ stabilizes the continuation path by reducing large wavefunction-parameter excursions, especially when fewer continuation steps are used and the FANPT updates must cover larger intervals in $\lambda$. 
This stabilization becomes important for both the CCSD(0) and CCSDT(2)Q(0) wavefunctions, and it is most visible in the worst-case same-$\lambda$ norm behavior.
\subsection{Computational Scaling and Continuation Quality}

Let $P$ denote the number of active wavefunction parameters, $M$ the number of determinants used to evaluate the wavefunction overlaps, and $Q$ the number of projected FANPT equations.

For qao$=2$, the constant vector contains contractions over a single parameter index, for example,

$$
B_n^{(r)}
\supset
\sum_{k=1}^{P}
G_{n,k\lambda}
p_k^{(r-1)}.
$$

For all $Q$ projected equations, this contraction requires

$$
\mathcal{O}(QP)
$$

operations per perturbation order.

The qao$=3$ approximation additionally requires the overlap Hessian,

$$
F_{m,kl}
=
\frac{\partial^2 f_m}
{\partial p_k \partial p_l},
$$

which has dimensions $M \times P \times P$. Its dense construction and storage therefore scale as

$$
\mathcal{O}(MP^2).
$$

The overlap Hessian is used to construct the three additional residual derivatives,

$$
G_{n,kl},
\qquad
G_{n,klE},
\qquad
G_{n,kl\lambda}.
$$

The tensor $G_{n,kl}$ requires a Hamiltonian application for every parameter pair $(k,l)$, while $G_{n,kl\lambda}$ requires an analogous application of the fluctuation potential. Their costs may therefore be written as

$$
\mathcal{O}(P^2 C_H)
\qquad \text{and} \qquad
\mathcal{O}(P^2 C_V),
$$

respectively, where $C_H$ and $C_V$ are the costs of applying the Hamiltonian and fluctuation-potential operators to one overlap vector. Since

$$
G_{n,klE} = -F_{n,kl},
$$

its construction requires

$$
\mathcal{O}(QP^2)
$$

operations.

The additional qao$=3$ constant-vector terms contain contractions over two parameter indices,

$$
B_n^{(r)}
\supset
\sum_{k,l=1}^{P}
T_{nkl}x_k y_l,
$$

and consequently scale as

$$
\mathcal{O}(QP^2).
$$

Thus, the order-dependent constant-vector construction changes from

$$
\mathcal{O}(QP)
\quad \text{for qao}=2
$$

to

$$
\mathcal{O}(QP^2)
\quad \text{for the additional qao}=3 \text{ terms}.
$$

The response matrix and the associated linear solve are identical for qao$=2$ and qao$=3$. Therefore, the incremental cost of qao$=3$ may be summarized as

$$
\Delta T
=
T_{\mathrm{qao3}}-T_{\mathrm{qao2}}
\sim
\mathcal{O}(MP^2)
+
\mathcal{O}(P^2 C_H)
+
\mathcal{O}(P^2 C_V)
+
\mathcal{O}(QP^2).
$$

Consequently, the current dense qao$=3$ implementation introduces one additional power of the number of active parameters relative to the qao$=2$ constant-vector construction. The same distinction appears in memory usage: qao$=2$ primarily stores parameter-indexed matrices, whereas qao$=3$ requires rank-three tensors with storage proportional to

$$
\mathcal{O}\left((M+Q)P^2\right).
$$

These estimates describe the present dense implementation. The prefactor and practical scaling may be reduced by exploiting the symmetry and sparsity of the overlap Hessian.
\begin{figure}[h]
\centering
\begin{subfigure}[b]{0.45\textwidth}
\centering
\includegraphics[scale=0.48]{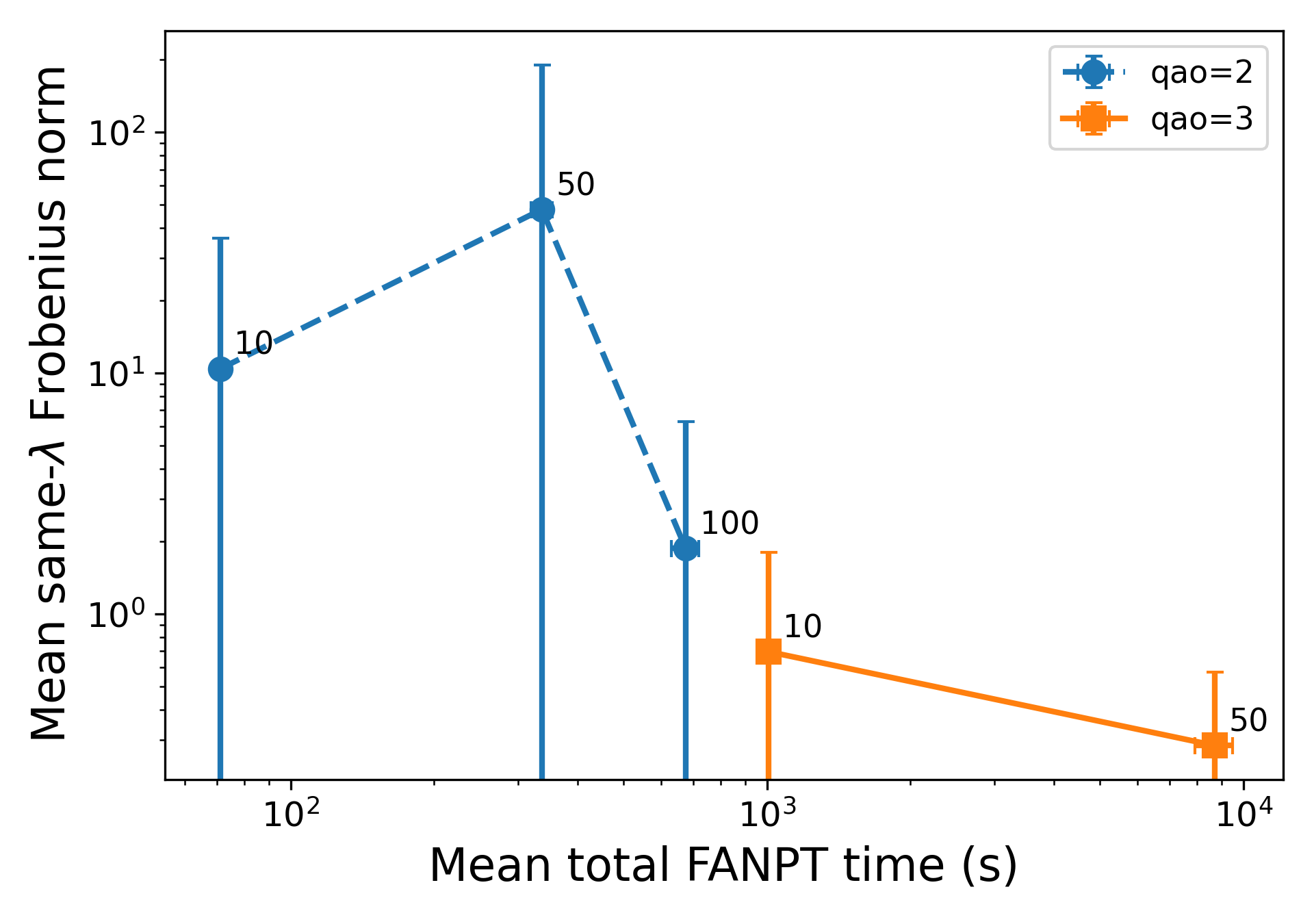}
\caption{}
\label{fig}
\end{subfigure}
\hfill
\begin{subfigure}[b]{0.45\textwidth}
\centering
\includegraphics[scale=0.48]{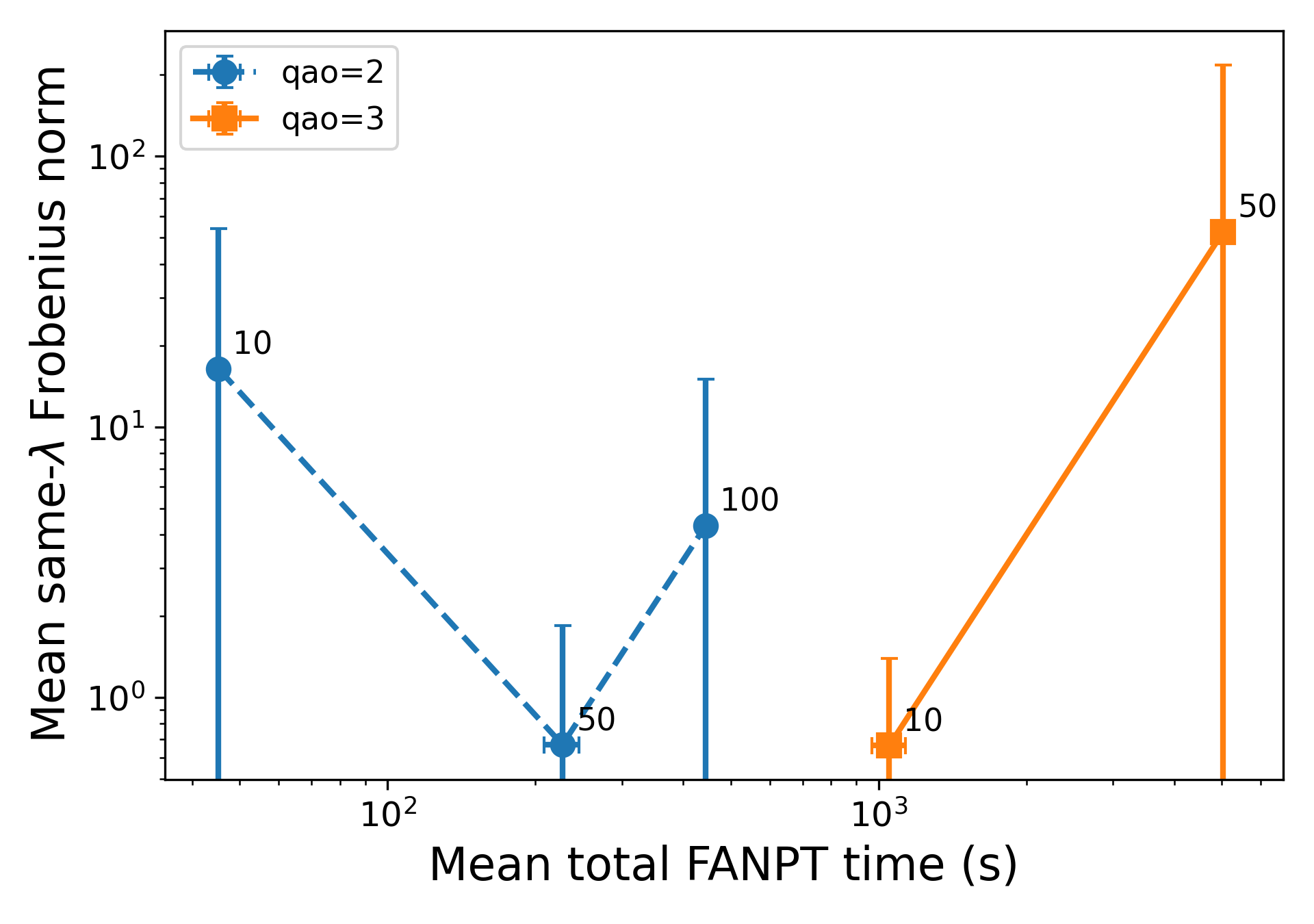}
\caption{}
\label{fig}
\end{subfigure}
\caption{Cost--quality comparison for the $C_{2v}$ insertion of \ce{Be} into \ce{H2} using the CCSDT(2)Q(0) wavefunction.
The mean same-$\lambda$ Frobenius norm of the parameter correction is plotted against the mean total FANPT wall time over the ten molecular geometries for (a) final order $2$ and (b) final order $3$. Labels indicate the number of FANPT steps, and error bars show one standard deviation across geometries. All calculations used the STO-6G basis set.}
\label{fig:fig_cost_quality}
\end{figure}

For the BeH$_2$ calculations (Figure~\ref{fig:fig_cost_quality}) with the CCSDT(2)Q(0) wave function, the empirical timings confirm that QAO$=3$ has a substantially larger FANPT cost than QAO$=2$; nevertheless, the higher-order approximation can provide markedly better continuation estimates using fewer steps. For final order $2$, QAO$=3$ with only $10$ steps required an average total FANPT time of approximately $1008$~s, compared with $675$~s for QAO$=2$ with $100$ steps. Although the QAO$=3$ calculation was about $1.5$ times more expensive in FANPT time, its mean same-$\lambda$ parameter norm decreased from $1.87$ to $0.70$, while the standard deviation across the ten geometries decreased from $4.39$ to $1.10$. Thus, the lower-step QAO$=3$ calculation reduced the average correction norm by approximately $63\%$ and its geometry-to-geometry variation by approximately $75\%$. A similar trend is observed for final order $3$: QAO$=3$ with $10$ steps produced a mean same-$\lambda$ norm of $0.665$ with a standard deviation of $0.733$, essentially matching the mean norm of $0.670$ obtained with QAO$=2$ and $50$ steps, while reducing the standard deviation from $1.18$. Relative to QAO$=2$ with $100$ steps, the same QAO$=3$, $10$-step calculation reduced the mean norm from $4.32$ to $0.665$ and the standard deviation from $10.69$ to $0.733$. For final order $3$, the large mean norm obtained with QAO$=3$ and $50$ steps is accompanied by a large standard deviation, indicating that the average is dominated by one or more unstable geometries rather than reflecting uniformly poor performance across the potential-energy curve. These results do not indicate that QAO$=3$ is intrinsically cheaper when considering the FANPT propagation alone; rather, they show that its additional computational cost produces a more accurate and, in several comparisons, substantially more uniform initial estimate across the potential-energy curve. This distinction is important when assessing practical utility: the relevant benefit of QAO$=3$ is not necessarily a reduction in the isolated FANPT wall time, but the possibility of reaching a given continuation quality with substantially fewer $\lambda$ steps and thereby reducing the difficulty and variability of the subsequent nonlinear FANCI optimizations.

\section{Conclusion}

In this work, we developed and implemented a second derivative corrected extension of FANPT for solving nonlinear FANCI wavefunction equations. The new approximation retains the overlap Hessian with respect to wavefunction parameters while neglecting only third and higher  parameter derivatives. This preserves the basic response-matrix structure of the original quasilinear FANPT formulation, but adds new constant-vector contributions involving the second derivatives of the projected residual. For linear CI wavefunctions, these additional terms vanish and the method reduces to the original FANPT approximation. For nonlinear ansatzes such as coupled cluster, however, the retained overlap-Hessian terms provide a more complete description of the local wavefunction response along the adiabatic connection.

Numerical tests on LiH and the BeH$_2$ insertion coordinate show that the second derivative corrected approximation does not substantially change the final FANPT energies, which are already very close to the independently optimized wavefunction energies. Its main benefit is instead the stabilization of the FANPT continuation path. For both CCSD(0) and CCSDT(2)Q(0) wavefunctions, especially when fewer $\lambda$ steps are used, qao $=3$ consistently reduces the largest same-$\lambda$ wavefunction-parameter deviations and suppresses severe parameter-space excursions observed with the original qao $=2$ approximation. These results indicate that the leading nonlinear overlap correction is most important not as an energy correction, but as a stabilizing term for robust continuation in nonlinear wavefunction calculations. This makes the second derivative corrected FANPT approximation a useful improvement for challenging systems where large perturbative steps or rapidly changing electronic structure can make the quasilinear trajectory unstable.

\section{Acknowledgement}
PWA thanks NSERC (ALLRP/592521-2023, RGPIN-2024-06707), the Canada Research Chairs (CRC-2022-00196), and the Digital Research Alliance of Canada. RAMQ acknowledges support from the National Science Foundation CAREER Award CHE-2439867.

\bibliography{referenceFile}

\end{document}